\documentclass[twocolumn,showpacs,preprintnumbers,amsmath,amssymb,aps,pre]{revtex4}
\usepackage{graphicx}
\begin{document}

\title{
Softening of Granular Packings with Dynamic Forcing
} 
\author{
C. J. Olson Reichhardt$^1$, L.M. Lopatina$^1$, X. Jia$^2$, and P.A. Johnson$^1$}
\affiliation{
$^1$Theoretical Division,
Los Alamos National Laboratory, Los Alamos, New Mexico 87545, USA\\
$^2$Institut Langevin, ESPCI ParisTech, CNRS UMR 7587 - 1 rue Jussieu, 
75005 Paris, France, EU
} 

\date{\today}
\begin{abstract}
We perform numerical simulations of a two-dimensional bidisperse granular
packing subjected to both a static confining pressure and a sinusoidal
dynamic forcing applied by a wall on one edge of the packing.  We
measure the response experienced by a wall on the opposite edge of
the packing and obtain the resonant frequency of the packing as the
static or dynamic pressures are varied.  Under increasing static pressure,
the resonant frequency increases, indicating a velocity increase of
elastic waves propagating through the packing.  In contrast, 
when the dynamic amplitude
is increased for fixed static pressure, the resonant frequency decreases,
indicating a decrease in the wave velocity.
This occurs both for compressional and for shear
dynamic forcing, and is in agreement with experimental results.  We find that
the average contact number $Z_c$ at the resonant frequency decreases with
increasing dynamic amplitude, indicating that the 
elastic softening of the packing
is associated with a reduced number of grain-grain contacts through which 
the elastic waves can travel.  We image the excitations created in the packing
and show that there are localized disturbances or soft spots 
that 
become more prevalent
with increasing dynamic amplitude.
Our results are in agreement with
experiments on glass bead packings and earth materials such as sandstone
and granite,
and may be relevant to the decrease in elastic wave velocities
that has
been observed to occur near fault zones after strong earthquakes, in surficial sediments during strong ground motion,
and in structures during earthquake excitation.
\end{abstract}
\maketitle

\vskip2pc

\section{Introduction}

Granular media has very unusual properties and can exhibit liquidlike
behavior by flowing under certain excitations, while it can have a solidlike
resistance to shear for other excitations.  The jamming phase diagram,
originally proposed by Liu and Nagel \cite{LiuNagel}, provides a convenient
description of the transition from jammed to unjammed states as a function
of density, temperature, or loading.  
Granular media can exhibit fragile properties in which the response
depends on the loading history \cite{cates,2001ostrovsky,2011jia,reichreview}.
A number of studies have
focused on the loading axis by applying a shear to the granular packing
and studying the unjamming of the packing above a certain shear level
\cite{s1,s2,s3,s4,s5,s6,s7,s8,s9}.  
Much work has also been performed on calculating the
normal or soft modes of granular packings \cite{n1,n2,n3,n4}, with particular
emphasis on the emergence of low frequency modes close to the jamming
transition.  

Most previous studies of granular matter under shear loading have considered
primarily a single direction of shear, or quasistatic shearing
\cite{s1,s2,s3,s4,s5,s6,s7,s8,s9}.  Relatively little
work has been performed on ac excitation 
or oscillation of granular matter in the
dense state.
Such dynamic shearing
of dense granular packings is of particular interest in connection with prominent effects
in surficial sediments from strong ground shaking from earthquakes 
\cite{field,beresnev}
as well as
the behavior of fault gouge material in response to earthquake forcing
\cite{2005johnson,2008johnson}.  Gouge is
a disordered  granular matter that often exists along and within the fault plane; it is produced by the long-term 
grinding of the tectonic
plates against each other via a process known as communition \cite{gouge}.    It has been hypothesized to
play a role in unusual nonlinear behavior of earthquake faults,
such as a delayed triggering response in which a large distant earthquake
can initiate an earthquake after a waiting time of days or months
\cite{trigger}.  Moreover, large earthquakes have been observed to
cause a long-lived depression of the 
elastic wave velocity in the mid to upper crust in
localized areas, which slowly recovers over time 
\cite{Brengieur2008,Delorey} 
as well as in near surface sediments \cite{2004schaff,2007rubinstein,2011rubinstein}.  

Experiments performed
with glass bead packs \cite{2005johnson,2011jia} 
and on natural materials such as sandstone 
show a similar decrease in the elastic wave velocity under oscillatory
or dynamic loading \cite{johnsonPT,2001ostrovsky,2004tencate}.
One common method for probing the softening of the elastic wave velocity 
is the use of nonlinear resonant
ultrasound spectroscopy, which can measure the nonlinear elastic state of
a rock or a glass bead pack \cite{2001ostrovsky}.
The frequency of an applied wave of
fixed amplitude $A$ is
swept or stepped across a resonant mode of the sample, and the resulting signal
is measured on the opposite side of the sample \cite{johnsonPT}.
In diverse materials including Berea sandstone, Lavoux limestone,
or synthetic slate, the resonant frequency drops with
increasing amplitude of the driving wave $A$ 
\cite{johnsonPT,2000smith,2005johnson2,2011tencate},
and this
indicates a drop in the velocity at which an elastic wave pulse travels
through the sample \cite{1996johnson,2004mashinskii,2011jia,2013knuth}.
In granular media,
when the static confining pressure is increased, the elastic wave velocities
increase \cite{goddard90,99jia,2004makse,2007zimmer}.
Early work on elastic wave or sound propagation in a glass bead packing
suggested that
the detailed contact structure of grains within the packing play an
important role in wave transmission \cite{92Liu,93Liu}, particularly
in short-wavelength wave scattering \cite{99jia}.  For long-wavelength
coherent waves, effective medium theory indicates
a link between the coordination number (the average number of contacts
per grain) and the elastic wave velocity \cite{duffy,digby,goddard90,2004makse}.
Simulations and experiments with 3D packings indicated that
the effective medium theory fails to account quantitatively for the shear
elastic modulus when the affine approximation breaks down at low static
pressures or high dynamical amplitudes \cite{makse99,agnolin,2013vdw}.
Much is understood regarding grain behavior under shear
\cite{ferdowsi13,ferdowsi14}; however,
despite a number of studies on sound wave propagation in two and three
dimensional packings \cite{2005somfai,2010khidas}, a
detailed microscopic understanding of the 
elastic wave velocity
evolution with driving amplitude
has not yet been obtained.

In this work, we study confined granular packings subjected to both a
static pressure and to dynamic loading achieved by applying an ac compressional
or shear loading to our model system.  We show that the behavior of the 
elastic wave propagation
matches what has been observed experimentally, and demonstrate that
changes in the contact number of the grains are correlated with the 
elastic wave propagation
changes.  We illustrate the dynamical motion of the grains and
discuss the implications of our work to dynamical triggering studies 
performed using earthquake catalogs.

\begin{figure}
\includegraphics[angle=-90,width=3.5in]{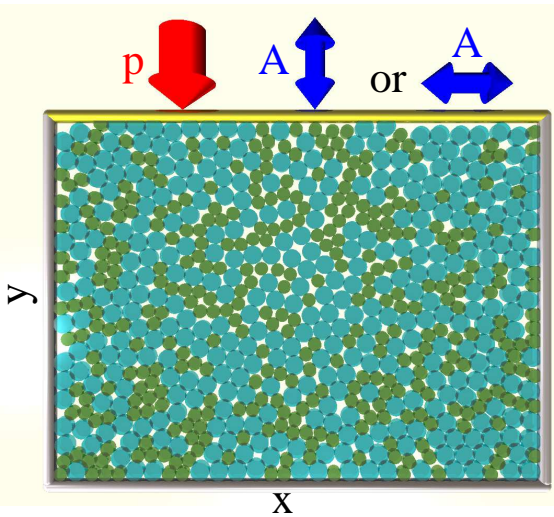}
\caption{Schematic of system showing the four confining walls.  The bottom and
side walls (grey) are fixed, while the top wall (yellow) 
is subjected both to a static 
confining force ${\bf F}_s^l=-p{\bf \hat y}$
(thick red arrow) and a sinusoidal dynamic loading force 
(thin blue arrows) in either the compressional (center arrow) or shear
(right arrow) direction.}
\label{fig:schematic}
\end{figure}

\section{Simulation}   

We consider a two-dimensional (2D) packing of 
$N=700$ disks with Hertzian contact
interactions \cite{hertz,cundall}:
\begin{equation}
{\bf F}_{ij}^{gg}=g\left[ \frac{1}{2}(D_i + D_j) - r_{ij}\right]^{3/2}{\bf \hat r}_{ij}
\end{equation}
where $g=10$ is the elastic constant of the grains
in dimensionless units, $D_{i(j)}$ is the diameter of
particle $i(j)$, ${\bf r}_{ij}={\bf R}_i-{\bf R}_j$, $r_{ij}=|{\bf r}_{ij}|$,
and
${\bf \hat r}_{ij}=({\bf R}_i-{\bf R}_j)/r_{ij}$.
The two grains interact only when they are in contact with
each other, for $r_{ij}\leq (D_i+D_j)/2$.  To avoid crystallization
of the packing, we use a bidisperse assembly of grains consisting of a 
50:50 mixture of grains with a radius ratio of 1:1.4.
We measure length in units of $a_0$, the diameter of the smaller of the
two sizes of grains.
We include shear friction between the grains \cite{gallas} of the form:
\begin{equation}
{\bf F}_{ij}^n=-\gamma_n m_{\rm eff}({\bf r}_{ij} \cdot {\bf v}_{ij}){\bf \hat r}_{ij}
\end{equation}
for the normal friction and
\begin{equation}
{\bf F}_{ij}^t=-\gamma_s m_{\rm eff}({\bf t}_{ij} \cdot {\bf v}_{ij}){\bf \hat t}_{ij}
\end{equation}
for the tangential friction.  Here $\gamma_n=0.1$ and $\gamma_s=0.1$ are the
dissipation coefficients, $m_{\rm eff}$ is the effective mass of the two grain
system, ${\bf v}_{i(j)}$ is the velocity of grain $i (j)$,
${\bf v}_{ij}={\bf v}_i-{\bf v}_j$, and
\begin{equation}
{\bf t}_{ij}=\left( \begin{array}{c}
-r_{ij}^y \\
r_{ij}^x \end{array}
\right).
\end{equation}
We employ a granular dynamics simulation technique to integrate the equations
of motion for each particle, given by
\begin{equation}
M_i{\bf \ddot r}_i = \sum_j \delta {\bf F}_{ij}^{gg}
\end{equation}
and
\begin{equation}
I_i {\ddot \phi}_i=\sum_j \delta C_{ij}
\end{equation}
where 
$M_i (I_i)$ is the mass (radius of gyration) of grain $i$, 
$\phi_i$ is the angular degree of freedom of grain $i$, and
$\delta C_{ij}$ is the torque exerted on a grain through contact with
other grains.

The grains are confined within four walls in our simulation box
as illustrated in Fig.~\ref{fig:schematic}, with no
periodic boundary conditions.  Wall interactions are modeled using image
grains that are the reflection of a grain in contact with the wall to the
other side of the wall.
The bottom and side walls are held at fixed
positions, while the top wall is a piston used to apply a static load 
${\bf F}_l^s=-p{\bf \hat{y}}$
normal to the wall
modulated by a dynamic load of the form 
${\bf F}_l^d=A \sin{\omega t}{\bf \hat{\alpha}}$,
where $\alpha=y$ for compressional loading and $\alpha=x$ for
shear loading.  
The position of the top wall is allowed to vary according to the
sum of the total load ${\bf F}_{\rm load}={\bf F}_l^s+{\bf F}_l^d$ and
the effective forces exerted on the image grains by the actual grains.
A static pressure value of $p=0.005$ corresponds to a downward force
on individual grains touching the top wall of $1.92\times 10^{-4}$.
This pressure is transmitted throughout the packing and opposed by an
effective force arising from the fixed bottom wall, so that individual
grains move very little when the static pressure is modified.
Similarly, a dynamic compressional amplitude of $A=0.030$ contributes
an oscillating force of magnitude $1.15\times 10^{-4}$ to each grain
touching the top wall, such that the motion of individual grains remains
much smaller than $a_0$.
It is important to note that 
due to the confinement, once the grains have been prepared
in the packing they are not able to rearrange their positions but can
only make slight shifts relative to their neighbors, which do not change.
We measure the net force exerted by the grains on the top wall, ${\bf f}^t(t)$,
and the bottom wall, ${\bf f}^b(t)$, for fixed $A$ while 
slowly stepping $\omega$ across a resonant frequency $\omega_0$.
For each driving frequency
$\omega$, we collect data during a period of 50 drive
cycles.
We then compute the power spectrum $S(\nu)$ of both
${\bf f}^t(t)$ and ${\bf f}^b(t)$, and obtain the
response in the form of the relative 
or normalized amplitudes of the output to input
signals at the driving frequency,
$\eta(A)=S(\nu=\omega/2\pi)_b/S(\nu=\omega/2\pi)_t$.

\begin{figure}
\includegraphics[width=0.5\textwidth]{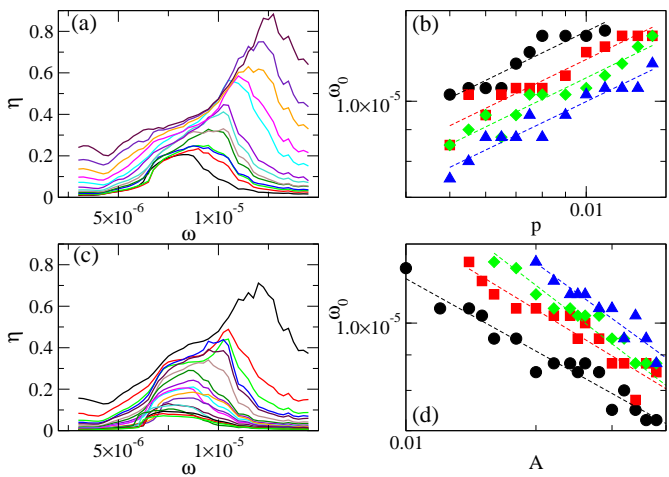}
\caption{
Results from the compressional dynamic simulation.
(a) Scaled amplitude of detected response 
$\eta$ vs driving frequency $\omega$
at $A=0.025$
for increasing static pressure $p=0.0050$, 
0.0055, 0.0060, 0.0065, 0.0070, 0.0075, 0.0080,
0.0090, 0.0100, 0.0110, 0.0120, 0.0130, and 0.0140
(bottom to top), showing a shift of the
resonant peak $\omega_0$ to higher frequencies with increasing $p$.
(b) 
Resonant frequency $\omega_0$ vs static pressure $p$
on a log-log scale, indicating an increase
in the elastic wave velocity with increasing static pressure, for
different values of the dynamic amplitude $A=0.015$, 0.020, 0.025,
and 0.030, from top to bottom.
Dashed lines are fits to $\omega_0 \propto p^\beta$ with $\beta\approx 0.35$.
(c)
$\eta$ vs $\omega$
at $p=0.0050$
for increasing dynamic amplitude $A=0.010$, 
0.012, 0.014, 0.015, 0.016, 0.018, 0.020, 0.022, 0.024,
0.025, 0.026, 0.028, 0.030, 0.032, 0.034, 0.036, 0.038, and 0.040
(top to bottom), showing a shift of the
resonant peak to lower frequencies with increasing $A$.
(d)
Resonant frequency $\omega_0$ vs dynamic amplitude $A$, 
on a log-log scale, indicating a decrease
in the elastic wave velocity with increasing $A$, for
different values of the static pressure $p=0.005$, 0.007, 0.009, and
0.011, from bottom to top.
Dashed lines are fits to $\omega_0 \propto A^{-\beta}$ with $\beta\approx 0.4$.
}
\label{fig:compress1}
\end{figure}

To prepare our
system, we remove the left wall of the sample, hold the top or piston
wall in a fixed position, and fill the system with
a granular gas.  We add a gravitational force term
${\bf F}_g=m_ig_g{\bf \hat x}$ to each grain and
allow the grains to settle into a dense packing.  We then close the left wall
and change the gravitational force to
${\bf F}_g=-m_ig_g{\bf \hat y}$ 
to force the grains toward the bottom wall of the packing;
we then permit the piston or top wall to move and incrementally apply a
static pressure to the piston, allowing the granular arrangement to settle to
a state of no net motion between pressure increments.  Once we have reached
the desired static pressure level $p$, we add a sinusoidal term to the force
exerted by the piston, resulting in a sinusoidal motion of the piston.
We permit the system to oscillate for 20 cycles in order to eliminate
any transient effects, and then measure the wall forces ${\bf f}^t(t)$
and ${\bf f}^b(t)$
during a period of 50 cycles.  In a given run we perform a frequency
sweep by holding the amplitude of the oscillation of the piston fixed
but increasing the frequency of the oscillation to a new value
after each set of 70
cycles.  To change the static pressure or the magnitude of the dynamic
forcing, we start with a fresh uncompacted sample in each case.  This
avoids a systematic increase in density that could otherwise occur after
each frequency sweep.
Our simulation measurement protocol is similar to that used in the
experiment described in
Ref.~\cite{2005johnson}, where the resonance compressional P waves
are observed.

\begin{figure}
\includegraphics[width=0.5\textwidth]{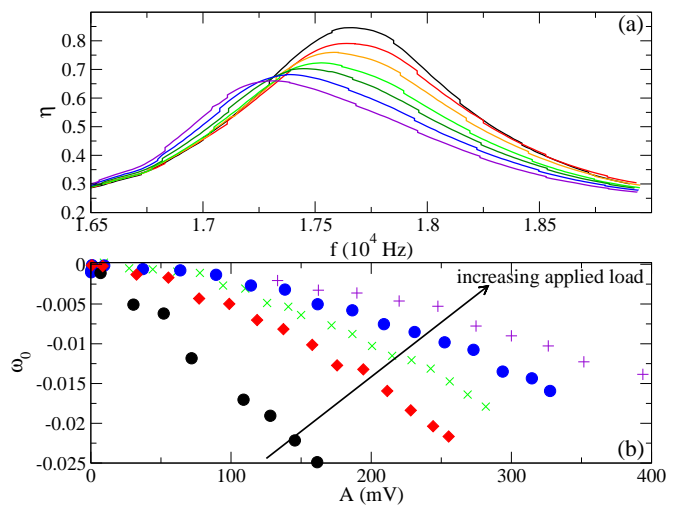}
\caption{
Experimental glass bead pack results modified from Ref.~\cite{2005johnson}.
(a) $\eta$ vs $f=\omega/2\pi$ for 
increasing dynamical amplitude $A = 10$ mV, 70 mV, 130 mV, 190mV,
250 mV, 310 mV, and 370 mV, from top to bottom.
(b)
Normalized $\Delta\omega_0$ vs $A$, in mV, for samples with increasing $p$ from
bottom to top.  The elastic wave velocity decreases with increasing $A$
in each case, but the overall magnitude of the decrease becomes smaller
as $p$ increases.
}
\label{fig:exp}
\end{figure}

\section{Results}

\begin{figure}
\includegraphics[width=0.5\textwidth]{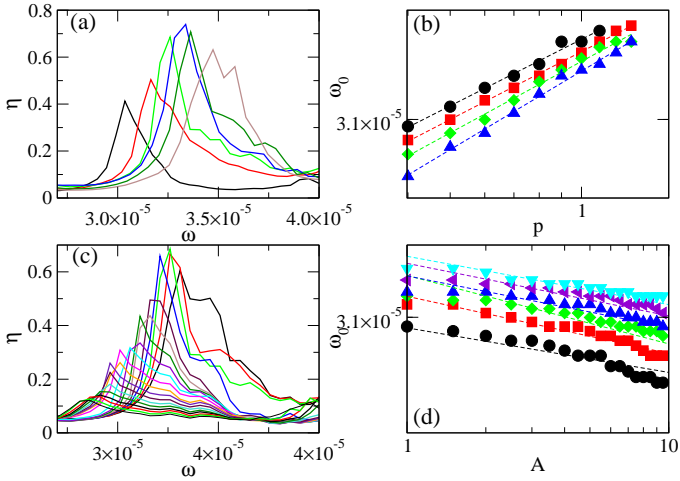}
\caption{
Results from the shear dynamic simulation.
(a)
$\eta$ vs $\omega$ at $A=2.0$ for increasing static
pressure $p=0.3$, 0.4, 0.5, 0.6, 0.7, and 0.8, from left maximum to right
maximum, showing a shift in $\omega_0$ to higher frequencies with increasing
$p$.
(b)
Resonant frequency $\omega_0$ vs static pressure $p$ on a log-log scale,
indicating an increase
in the elastic wave velocity with increasing static pressure, for
different values of the dynamic amplitude $A=3$, 5, 7, and 10, from
top to bottom.
Dashed lines are fits to $\omega_0 \propto p^\beta$ with $\beta \approx 0.25$.
(c)
$\eta$ vs $\omega$ at $p=0.5$ for increasing dynamic amplitude $A=1.0$, 1.5,
2.0, 2.5, 3.0, 3.5, 4.0, 4.5, 5.0, 5.5, 6.0, 6.5, 7.0, 7.5, 8.0, 8.5, 9.0,
9.5, and 10.0, from left maximum to right maximum, showing a shift of 
$\omega_0$ to lower frequencies with increasing $A$.
(d)
Resonant frequency $\omega_0$ vs dynamic amplitude $A$, on a log-log scale,
indicating an increase
in the elastic wave velocity with increasing $A$, for
different values of the static pressure $p=0.3$, 0.4, 0.5, 0.6, 0.7, and
0.8, from top to bottom.  Dashed lines are fits to $\omega_0 \propto A^{-\beta}$
with $\beta \approx 0.1$.
}
\label{fig:shear1}
\end{figure}

We first compare measurements of the dynamic response $\eta$ in
the compressional and shear oscillatory simulations and in experiments.
In Fig.~\ref{fig:compress1}(a) we plot the normalized amplitude $\eta$ 
as a function of
driving frequency $\omega$ for fixed 
compressional dynamic amplitude $A=0.025$ and
static pressures ranging from $p=0.005$ to $p=0.0140$.
Here the resonant frequency $\omega_0$ shifts to higher values as the
static pressure is increased.  This indicates that the elastic wave 
velocity is
increasing with increasing static pressure, in agreement with
previous observations.
By identifying the value of $\omega_0$ from each curve, we construct a plot
of $\omega_0$ versus $p$ shown in Fig.~\ref{fig:compress1}(b) 
for dynamic amplitudes
ranging from $A=0.015$ to $A=0.030$.
The resonant frequency increases with increasing
static pressure roughly as a power law with slope $\beta\approx 0.35$;
however, there is an overall downward shift in the
resonant frequency as the 
compressional dynamic loading $A$ increases.
In Fig.~\ref{fig:compress1}(c) we plot $\eta$ versus $\omega$ 
for the compressed system
at fixed static pressure $p=0.0050$ and dynamic amplitudes ranging
from $A=0.010$ to $A=0.040$.
Here, the peak value $\omega_0$ {\it decreases} in frequency with increasing
dynamic amplitude $A$, indicating that the elastic wave velocity is
{\it decreasing} with increased dynamic driving.  This softening of
the system with dynamic driving is more clearly shown in 
Fig.~\ref{fig:compress1}(d) where
we plot $\omega_0$ versus $A$ for values of $p$ ranging from 
$p=0.005$ to $p=0.011$.  The softening is very robust and appears for
each value of $p$.
For comparison, we illustrate in Fig.~\ref{fig:exp}(a) the
experimentally obtained values of $\eta$ as a function of frequency for
different dynamical amplitudes $A$.  The resonant frequency
decreases with increasing dynamic amplitude.  This is more clearly
shown in Fig.~\ref{fig:exp}(b), where we plot $\Delta\omega_0$,
the shift in $\omega_0$ from a reference value, 
versus the dynamic amplitude $A$ for different values of static
pressure \cite{2005johnson}.  In each case, the resonant frequency decreases
with increasing dynamic amplitude in agreement with the simulation results.

We find similar behavior for a system in which the top plate is dynamically
sheared in the direction transverse to the applied static pressure.  
In Fig.~\ref{fig:shear1}(a) 
we illustrate representative $\eta$ vs $\omega$ curves at $A=2.0$ and 
increasing static pressure $p$ in the sheared system.  
Figure \ref{fig:shear1}(b) shows 
a log-log plot $\omega_0$ versus $p$ curves for values
of $A$ ranging from 3 to 10 in the same system.  
We observe a power law behavior $\omega_0 \propto p^\beta$ with 
$\beta \approx 0.25$, a somewhat smaller exponent than in the 
dynamically compressed
system.
The resonant frequency
increases with increasing $p$, indicating an increase in 
the elastic wave velocity with
increasing static pressure.  We note that significantly larger 
static pressures
must be applied to the dynamically 
sheared system than to the dynamically compressed system in order
to obtain a wave signal that propagates through the entire packing and is
measurable on the bottom plate.  
For increasing dynamic amplitude, the resonant frequency decreases,
as illustrated in Fig.~\ref{fig:shear1}(c) for $p=0.5$ and a range 
of values of $A$.
The decrease is slower than linear, as shown in 
Fig.~\ref{fig:shear1}(d) where we plot
$\omega_0$ versus $A$ for different values of $p$ in the 
dynamically sheared system.  These simulation results are also in
excellent agreement with our experimental observations on
shear resonant modes \cite{inprep}.

\begin{figure}
\includegraphics[width=0.5\textwidth]{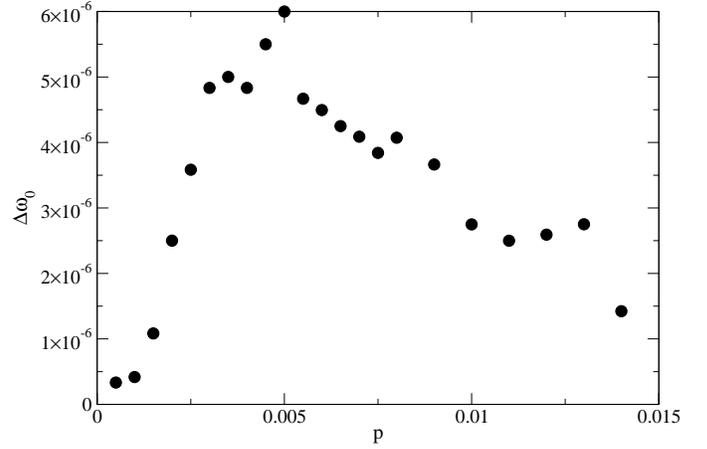}
\caption{ 
Results from the compressional dynamic simulation.
The total magnitude of the frequency shift
across our measured range of $A$, 
$\Delta \omega=\omega_0(A=0.005)-\omega_0(A=0.030)$ 
vs $p$ shows 
two regimes of frequency shift behavior.
At low $p$, $\Delta \omega_0$ increases
with increasing static pressure, while for $p>0.005$, 
$\Delta \omega_0$
decreases with increasing $p$.
The higher static pressure regime agrees with the experimental response.
}
\label{fig:tworegimes}
\end{figure}

In Fig.~\ref{fig:exp}(b) we find that 
experimentally, the overall magnitude of the decrease in $f_0$ with
increasing dynamic amplitude, $\Delta \omega_0$, 
becomes smaller when the static load $p$ is
increased.  The same behavior occurs in the simulations, as shown in
Figs.~\ref{fig:compress1}(d) and \ref{fig:shear1}(d). 
For the compressional dynamic simulations, we find that 
if we decrease the static pressure $p$ to very small values, 
$\Delta \omega_0$ passes through a peak value and then begins to
decrease with decreasing $p$ instead of increasing.
This is shown in Fig.~\ref{fig:tworegimes},
where we plot 
$\Delta \omega_0=\omega_0(A=0.005)-\omega_0(A=0.030)$ 
as a function of
static pressure $p$ in the compressional system.  For $p<0.005$,
$\Delta \omega_0$ increases with increasing static pressure, while for
$p>0.005$, $\Delta \omega_0$ decreases with increasing static pressure. 
The higher $p$ behavior 
agrees with the experimental results
\cite{2005johnson}.  For the remainder of this
paper, we will focus on the higher pressure regime with $p>0.005$ in
the compressional dynamic simulation.

\begin{figure}
\includegraphics[width=0.5\textwidth]{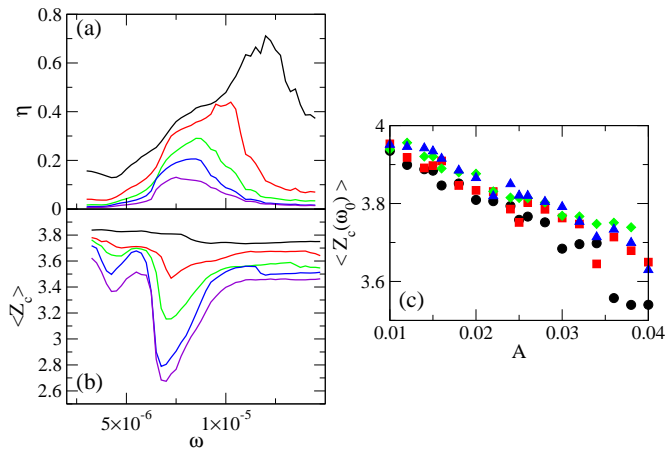}
\caption{
Results from the compressional dynamic simulation.
(a) $\eta$ vs $\omega$
for $A=0.010$, 0.015, 0.020, 0.025, and 0.030, from top
to bottom, at fixed $p=0.005$.
(b) Average contact number $Z_c$ in the packing vs $\omega$
for $A=0.010$, 0.015, 0.020, 0.025, and 0.030, from
top to bottom, at fixed $p=0.005$.
There is a pronounced dip
in $Z_c$ that increases in magnitude with increasing $A$. 
(c)
Value of $\langle Z_c\rangle$ at $\omega=\omega_0$ as a
function of dynamic amplitude $A$ for 
$p=0.005$, 0.007, 0.009, and 0.011, from bottom to top.
}
\label{fig:zavg}
\end{figure}

We next compare the response of the system with the average coordination
number $\langle Z_c\rangle=N^{-1}\sum Z_i$ 
of the packing, where $Z_i$ is the number of
particles in direct contact with particle $i$.  In Fig.~\ref{fig:zavg}(a) 
we plot
the normalized amplitude versus driving frequency in a compressional
system with $p=0.005$ for dynamic amplitudes 
ranging from $A=0.010$ to $A=0.030$.
As before, we observe that the resonant frequency $\omega_0$ decreases with
increased $A$.  In Fig.~\ref{fig:zavg}(b) we show the corresponding
$\langle Z_c\rangle$ versus driving frequency.  
Near the resonance frequency $\omega \approx 7\times 10^{-6}$ there is a
dip in $\langle Z_c\rangle$ 
which increases in magnitude with increasing $A$,
indicating that the packing is becoming looser as the dynamic amplitude
increases.
The decrease of $\langle Z_c(\omega_0)\rangle$, 
the value of $\langle Z_c\rangle$ at the
resonant frequency, is shown in Fig.~\ref{fig:zavg}(c) 
as a function of $A$.
There is
a slight increase in 
$\langle Z_c(\omega_0)\rangle$ as the static pressure increases, but there
is a clear decrease in 
$\langle Z_c(\omega_0)\rangle$ with increasing $A$.  
As suggested by the effective medium theory, the elastic wave velocity
is proportional to the coordination number
\cite{duffy,digby,goddard90,2004makse}.
Thus the reduction of
the number of the contacts in the packing is the physical reason for the
decrease in the elastic wave velocity with increasing dynamical amplitude in the
granular packing.  
Higher static pressure
forces more grains into direct contact.
In contrast, larger amplitudes of dynamical
forcing tend to break contacts in the packing.

\begin{figure}
\includegraphics[width=0.5\textwidth]{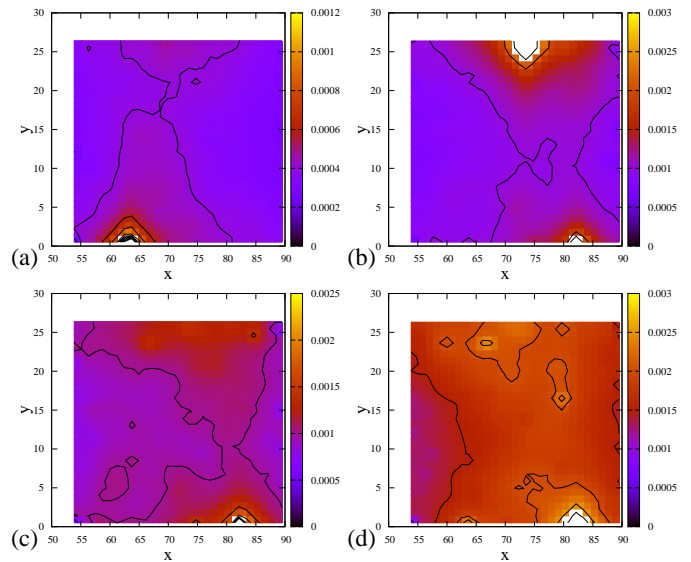}
\caption{Contour plot showing the
regions of the sample undergoing the largest amount
of motion in the dynamically compressed system at $p=0.005$
at the resonant frequency.
Colors indicate the magnitude of the motion at each spatial location.
(a) $A=0.010$. (b) $A=0.020$. (c) $A=0.025$. (d) $A=0.030$.}
\label{fig:3d}
\end{figure}

Finally, we find that
the motion of the grains under excitation takes two forms.  The bulk of the
packing responds collectively, with a large section of the packing moving
coherently in response to the dynamic forcing at and near resonance.  We
also observe isolated soft spots or rattler areas where an individual
grain has a much higher amplitude of motion than the grains that surround 
it.  These soft spots tend to contribute additional damping to the
propagating elastic wave signature.  To identify the 
soft spots
in the compressional dynamic simulation,
we compute $\delta r_i = \max({\bf r}_i(t)-{\bf r}_i(0))$, 
which is the maximum displacement 
of an individual particle from its average equilibrium position
${\bf r}_i(0)$ for a given driving frequency, amplitude, and static pressure.  
In Fig.~\ref{fig:3d} we show contour plots of the value of $\delta r_i$
at the resonant frequency for packings with $p=0.005$ and $A=0.010$ to
0.030.
The position in the packing is indicated on the $x$ and
$y$ axes, 
while the coloring indicates the value of $\delta r_i$.
The number and density of soft spots, indicated by local maxima
in $\delta r_i$, increases with increasing
driving amplitude, with a single spot in Fig.~\ref{fig:3d}(a),
two in Fig.~\ref{fig:3d}(b), three in Fig.~\ref{fig:3d}(c), and
more than four in Fig.~\ref{fig:3d}(d).
This proliferation of soft spots contributes to the drop in $\omega_0$
with increasing driving amplitude.

\section{DISCUSSION}

Different regimes of the elastic wave velocity $c$ 
(compressional or shear) through 2D or 3D bead packings
as a function of 
applied static pressure $p_{\rm ext}$ 
have been observed previously in experiment, with
$c \propto p^{1/6}$ at high pressures but $c \propto p^{1/4}$ at low pressure
\cite{duffy,goddard90,gilles,jia01,velicky02,gilles03,2008coste,2008brunet}.  
The results shown in Figs.~\ref{fig:compress1} and \ref{fig:shear1} are
more consistent with the low pressure regime.
For the case of monodisperse disk
packings, this effect was treated analytically in Ref.~\cite{2009pride}.
This scaling behavior might be correlated to the change in static
pressure with the average contact number in the packing
\cite{goddard90,henkes} as has been
confirmed numerically \cite{2004makse,2005somfai,2007majmudar} and
in experiments \cite{2010zhang}.

Regarding the magnitude of
the change in the wave velocity $c$ with dynamical amplitude, it is 
generally larger for low
static pressure than for high static pressure 
\cite{2004mashinskii,2005johnson}, in
agreement with our results in Figs.~\ref{fig:compress1} and \ref{fig:shear1}.  
In Ref.~\cite{2013vdw}, this
behavior was suggested to result from significant rearrangements of the
contact network, resulting in a change in the average contact number but 
without significant motion of particles or a significant change in
the packing density.  Indeed, as we observe here, small shifts in the
positions 
of individual grains can modify
the local contact number enough to change the effective velocity of
elastic waves in the system.
We find a reduction in the average contact number
when the amplitude of
the dynamical forcing is increased, consistent with the
experimental results.  This resembles the ``acoustic
fluidization'' effect (initially introduced for describing frictional
weakening \cite{79melosh}) that has been observed in which the
elastic wave velocity
can soften under large wave amplitudes even when significant contact
reorganization and sliding do not occur 
\cite{2011jia,2012espindola}.  If the
wave amplitude were large enough to drop the average contact number below
the jamming threshold in a significant portion of the sample, a ``sonic
vacuum'' state could occur in which transmission of elastic waves would
become impossible \cite{2012gomez}.  Above these amplitudes, the entire
packing fluidizes, as in Refs.~\cite{1994luding,1995luding}.

Granular packings often exhibit heterogeneous responses due to their
highly disordered internal contact structure.  In a 2D idealized
granular packing, based on the response of a single grain to a
sinusoidal driving frequency, localized normal modes at 
high frequency were predicted
to occur, likely due to the interference between scattered plane waves
\cite{1994leibig}.
Evidence for localized soft
spots has been observed in Hertzian packings where the velocity
distribution functions for the motion of individual particles have
fat tails, indicating strongly non-Gaussian behavior \cite{2013owens}.
These soft spots found at relatively low frequencies have been
connected particularly with highly nonlinear responses
such as glass-like behavior and non-affine displacement fields
in granular packings \cite{n1,n2,n3,n4,tsamados,manning,chen}.

Numerous studies have employed granular packings as a surrogate for the
complex behavior occurring along fault zones in Earth 
\cite{gouge,2005johnson,2008daniels,2008johnson,2012johnson}.
Our results may suggest that the decrease in velocity observed along and near
a fault after a large earthquake is analogous to a change in the
granular packing to a state with a reduced number of contacts, even if
no significant rearrangements of the grain positions have occurred. 
These contacts could gradually reconnect over time, in analogy with the
slow recovery of the sound velocity that has been observed in the earth.

Indeed, in the earth, 
the fault blocks surrounding a fault zone contain fractures at many scales.  These are analogous to the 
grain contacts in our simulation and laboratory experiments.  As wave amplitudes increase,
slip is mobilized along the fractures resulting in a bulk modulus softening of the rock.  
This behavior is followed by slow dynamics where contacts in fractures are re-established, as demonstrated in laboratory experiments
\cite{2005johnson,2011jia}.

\section{SUMMARY}
We characterize the evolution of the internal characteristics of 
bidisperse two-dimensional
granular packings under large amplitude dynamic forcing and 
varied confining pressure.  We find that the resonant frequency or
fundamental mode of the frequency decreases with increasing dynamic
amplitude at constant static pressure, in agreement with laboratory and field
experiments.  For fixed dynamic amplitude, the resonant frequency
increases with increasing confining pressure, also in agreement with 
experiment.  We show that the average contact number $Z_c$ of the packing
decreases both at resonance and for increasing dynamic amplitude.  We 
characterize the heterogeneity of the packing response by measuring the
vibration displacement of each grain, and find regions of high and low 
displacements.  Our approach provides insight into the elastic nonlinear
nature of unconsolidated materials such as granular packings, as well as
consolidated materials such as sandstone.

\acknowledgments
This work was supported by Institutional Support (LDRD) at Los Alamos National Laboratory.
This work was carried out under the auspices of the 
NNSA of the 
U.S. DoE
at 
LANL
under Contract No.
DE-AC52-06NA25396.

\end{document}